\documentstyle[aps,epsf,prb]{revtex}

\begin{document}
\draft

\twocolumn[\hsize\textwidth\columnwidth\hsize\csname@twocolumnfalse\endcsname
 
\title{Anomalous spin susceptibility and magnetic polaron formation in
the double exchange systems}

\author{ Hongsuk Yi and N. H. Hur}
\address{ Center for CMR Materials, Korea Research Institute of
  Standards and Science,\\ Yusong, P.O. Box 102, Taejon 305-600,
  Korea} 

\author{ Jaejun Yu} 
\address{ Department of Physics and Center for Strongly Correlated
  Materials Research,\\ Seoul National University, Seoul 151-742, Korea} 

\date{\today}

\maketitle
\begin{abstract} 
  
  The magnetic susceptibility and spin-spin correlation of the
  double-exchange model for doped manganites are investigated through
  the Monte Carlo calculations on the three-dimensional lattice model.
  Deviations of the susceptibility from the Curie-Weiss behavior above
  the ferromagnetic ordering temperature $T_c$ seem to indicate a
  formation of local ferromagnetic clusters in the vicinity of $T_c$,
  which is consistent with recent electron paramagnetic resonance
  experiments for La$_{2/3}$Ca$_{1/3}$MnO$_3$.  A further analysis of
  the spin-spin correlations show the ferromagnetic cluster size to be
  three-to-four lattice spacings, suggesting that the charge carriers
  may form magnetic polarons.

\end{abstract}
 ]
 \narrowtext

\section{Introduction}

Recent interests in mixed-valence manganites with the chemical
compositions of $R_{\rm 1-x}A_{\rm x}$MnO$_3$ (where $R$=rare earth;
$A$=Ca, Sr, Ba) are largely ascribed to their potential technological
applications.\cite{cmr} The most notable feature of these materials is
an extremely large change of resistivity under the application of a
magnetic field near ferromagnetic ordering temperature $T_c$, which is
known as colossal magnetoresistance (CMR). A metal-insulator (MI)
transition accompanied by the ferromagnetic-paramagnetic phase
transition is occurred near $T_c$. The strong connection between the
MI transition and the ferromagnetic spin alignment has been understood
in terms of the double-exchange (DE) mechanism.\cite{zener,kubo} The
conduction electrons in the $e_g$ orbital of Mn$^{3+}$ ions are
hopping in the background of Mn$^{4+}$ $(t_{2g})$ ion spins with an
experience of a strong on-site Hund's rule coupling, thereby leading
to an effective hopping integral of the form ${t}\cos(\theta_{ij}/2)$
where $\theta_{ij}$ is the relative angle between Mn$^{4+}$ ions.

Although the electron-lattice interaction arising from the dynamic
Jahn-Teller distortion is considered to be important for the
understanding of overall trends of CMR phenomena\cite{millis}, the
lattice polaron formation is incomplete to explain the transport
properties in connection with the observed CMR
phenomena.\cite{millis2,zang,jdlee} Contrary to the above approach, it
is suggested that the $e_g$ carriers can be trapped by spin-disorder
scattering due to the local deviations of the ferromagnetic
surroundings, resulting in the formation of magnetic polarons in the
vicinity of $T_c$.\cite{furukawa,varma} Indeed, a lot of theoretical
works\cite{horsch,batista,coey,hsyi,dagotto} have demonstrated that
the magnetic polaron formation plays a crucial role in CMR phenomena
in the paramagnetic state.

{}From the experimental point of view, much efforts have been devoted
to understand their unique magnetic features in the paramagnetic
state.  For instance, the high temperature inverse susceptibility for
La$_{0.7}$Ca$_{0.3}$MnO$_3$ is smaller than expected from the
Curie-Weiss law at low magnetic field $H < 0.1$~T.\cite{jaime,sun} The
deviation of susceptibility from Curie-Weiss law is also found in the
layered material of La$_{1.35}$Sr$_{1.65}$Mn$_2$O$_7$, suggesting the
formation of ferromagnetic cluster above $T_c$. \cite{chauvet} Recent
electron paramagnetic resonance (EPR) study\cite{oseroff} revealed
that EPR signals decrease exponentially at temperatures slightly above
$T_c$, which results in an enormous enhancement of the effective spins
of about $\sim 30$ per formula unit.  Moreover, small angle neutron
scattering experiment\cite{teresa} estimated the size of the
ferromagnetic cluster in the paramagnetic state to be about
$12$\AA. There has also been an increasing
realization\cite{radaelli,kajimoto,shengelaya} which manifest the
importance of the magnetic fluctuations that is beyond the mean-field
prediction in the paramagnetic state of the doped perovskite
manganites.

The main purpose of this paper is to elucidate the characteristic
features of the ferromagnetic polaron in the vicinity of $T_c$.  In
order to account for the spin fluctuation effects accurately, we adopt
a Monte Carlo method\cite{hsyi,dagotto} for the DE model on the
three-dimensional (3D) lattice. We found that the temperature
dependence of inverse susceptibility deviates from the expected
Curie-Weiss behavior, which is in good agreement with experimental
observations. Also, the activation energy estimated from the
non-Curie-Weiss part of the susceptibility coincides with that of EPR
measurements for La$_{2/3}$Ca$_{1/3}$MnO$_3$.\cite{oseroff} From these
results together with the calculated spin-spin correlation, we suggest
that charge carriers above $T_c$ form the ferromagnetic cluster with
the size of three-to-four lattice spacings.

\section{Model and Calculations}

One of the simplest models for the description of the
paramagnetic-ferromagnetic phase transition in doped manganites is a
single-orbital DE model Hamiltonian. In the strong Hund's coupling
limit ($J_H \rightarrow \infty$), it can be written
as\cite{furukawa,varma,horsch}
\begin{eqnarray}
\label{eq:ham}
{\cal H} = -\sum_{\langle ij \rangle } \Bigl( t_{ij} c_{i }^+ c_{j } +
{\rm h.c.}\Bigr) -h\sum_iS_i^z,
\end{eqnarray}
where the operator $c^+_{i}$ creates a spinless conduction electron at
site $\vec{R}_i$ and $h$ is an external magnetic field. The hopping
amplitude in the strong Hund's coupling limit is given by
\begin{eqnarray}
t_{ij}=t\Bigl(\cos\frac{\theta_i}{2}\cos\frac{\theta_j}{2} +
\sin\frac{\theta_i}{2}\sin\frac{\theta_j}{2}e^{i(\phi_i-\phi_j)}
\Bigr) \ ,
\end{eqnarray}
where the localized $t_{2g}$ spin $\vec{S_i}$ is treated as a
classical spin $\vec{S_i}=|\vec{S_i}|(\sin\theta_i\cos\phi_i \hat{x}+
\sin\theta_i\sin\phi_i\hat{y} + \cos\theta_i \hat{z})$ with the polar
angles $\{ \theta_{\bf i}, \phi_{\bf i} \}$ characterizing the
orientation of the localized spin $\vec{S}_i$ since the quantum
effects can be neglected near room temperature.  Furthermore, we
assume that the azimuthal angle $\phi_i$ rotates independently since
the contribution of the phase factor in the hopping integral to the
partition function can be neglected.\cite{hsyi,dagotto} Thus, it is
possible to replace $t_{ij}$ by the DE form $t_{ij} =
t\cos(\theta_{ij}/2)$, where $\theta_{ij}$ is the relative angle
between neighboring Mn ions, and the conduction electrons are treated
as a quantum object.

The grand canonical partition function of the present model with
chemical potential can be denoted by
\begin{eqnarray}
Z={\rm Tr}_{S}{\rm Tr}_{c}e^{-\beta({\cal H} - \mu {\cal N})},
\end{eqnarray}
where ${\rm Tr}_{S}$ and ${\rm Tr}_{c}$ represent traces over the
localized spin configuration and the conduction electron degree of
freedom, respectively, and ${\cal N}=\sum_{i}c^+_{i}c_{i}$.  Here the
conduction electron density $\langle n \rangle=\langle {\cal N}
\rangle/L^3$ is determined by adjusting chemical potential
$\mu$. Since the DE model Hamiltonian is quadratic with respect to the
Fermion operator $\{ c_i \}$, ${\rm Tr}_{c}$ is directly calculated by
diagonalizing the electronic part of the Hamiltonian.  By performing
the diagonalization of the $(L^3\times L^3)$ hermitian matrix for each
given spin configuration $\{\theta_{\bf i},\phi_{\bf i}\}$, we obtain
the $L^3$ eigenvalues denoted by $\epsilon_\alpha(\theta_{\bf
i},\phi_{\bf i})$.  Thus the resulting partition function becomes
\begin{equation}
Z = \prod^{L^3}_{\bf i} \left(\,\int^{\pi}_0 d\theta_{\bf i}
\sin\theta_{\bf i} \int^{2 \pi}_0 d\phi_{\bf
i}\,\right)\,\prod^{L^3}_{\alpha = 1} (1 + e^{-\beta
(\epsilon_{\alpha}-\mu)}).
\end{equation}
Now, we can apply a Monte Carlo integration procedure for the
summation over the configuration angles $\{ \theta_{\bf i}, \phi_{\bf
i} \}$ of localized spins using a standard Metropolis algorithm. The
actual calculations are performed for 3D cubic lattices $L^{3}=6^{3}$
with periodic boundary conditions in spatial directions.  Typically,
we take 5000 Monte Carlo steps per site for statistical average.
Thermodynamic quantities of interest are obtained directly from the
thermal average of spin configurations and the eigenvalues of
Hamiltonian. The carrier density $\langle n \rangle \approx 0.5$,
i.e., the hole density $x =1- \langle n \rangle $, obtained by fixing
$\mu=0.0$.

\section{Results}

Before studying the magnetic fluctuation effects, we first investigate
the nature of the magnetic transition in the ferromagnetic state.
Figure~1 shows the temperature $(T)$ dependence of the magnetization,
$M=\langle {\sum_i S_i^z} \rangle /L^3$, denoted by solid squares.  As
clearly shown in Fig.~1, the magnetization drops sharply when the
ferromagnetic ordering temperature reaches to $T_c\approx 0.125t$ and
the so-called ferromagnetic-to-paramagnetic transition takes places
near $T_c$. The dashed line represents the mean-field prediction of DE
model by Kubo and Ohata, leading to $M=M_o|T_c-T|^{1/2}$ with
$M_o=\sqrt{5/{T_c}}$ for $S=2$ as $T_c$ is approached.\cite{kubo} This
magnetic transition is closely related to the other transition of
apparently different character, i.e., the $T$-dependence of
resistivity.\cite{furukawa} The obtained Monte Carlo result, however,
varies more slowly than that of the mean-field prediction of DE
model. Moreover it is noted that the magnetization data can be fitted
with a power law of $M(T)\sim|T_c -T|^{\beta}$ just below $T_c$.  The
solid line in the Fig.~1 is a fitting curve, yielding $\beta=0.32 \pm
0.01$. This value of the exponent $\beta$ is very close to that of
La$_{0.7}$Sr$_{0.3}$MnO$_3$ single crystal\cite{ghosh} and 3D
Heisenberg class which is $\beta=0.33$ in the vicinity of $T_c$. Above
$T_c$, however, our calculated result of $M$ is different from zero
due to the finite-size effects. The similar results are also obtained
by using different Monte Carlo methods for a larger size of unit cells
$20\times 20 \times 20$.\cite{calderon}

\begin{figure}
\centering \leavevmode \epsfxsize=7.8cm \epsfbox{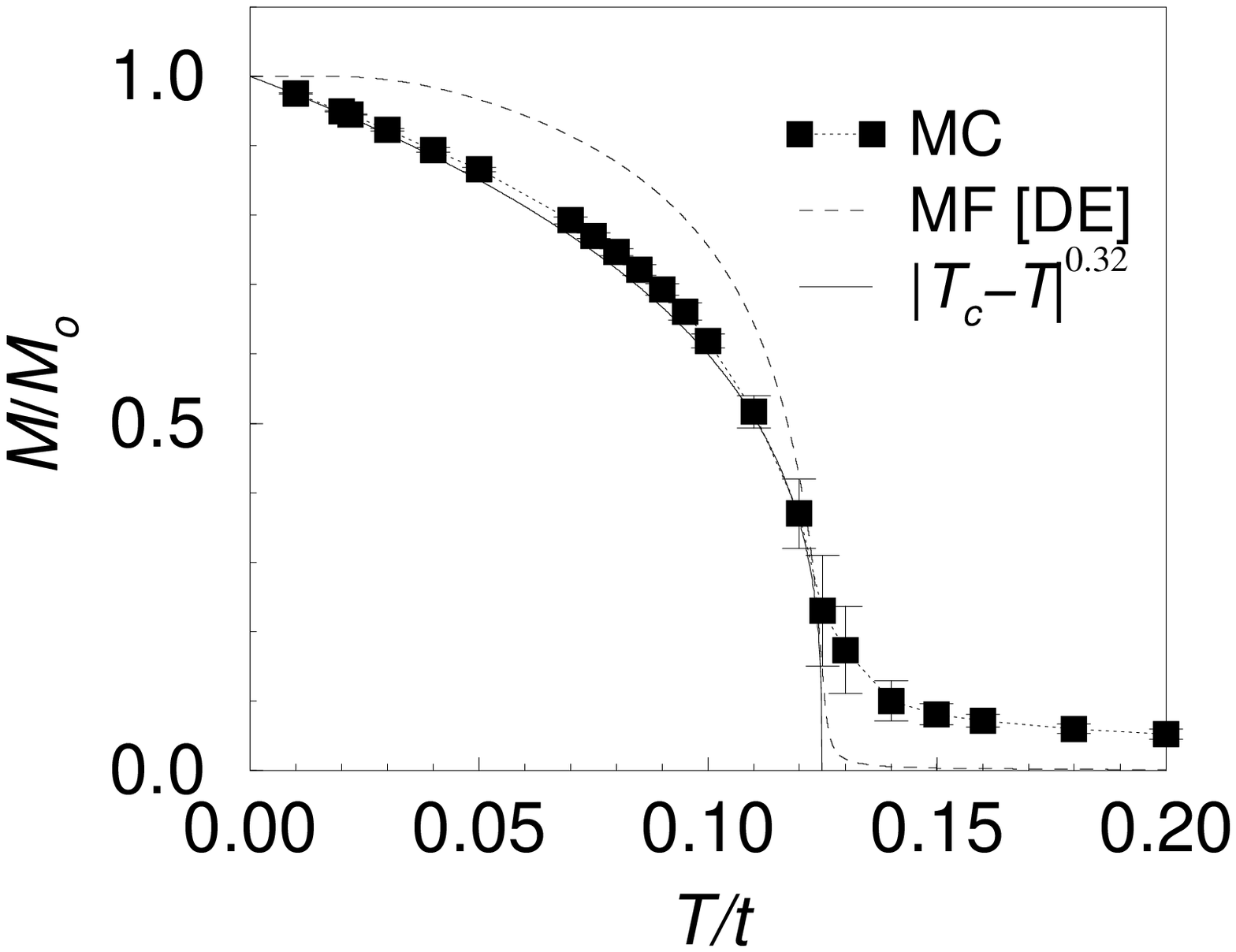}
\caption{Temperature dependence of magnetization $M(T)$. The dashed
and the solid line is the mean-field prediction of double-exchange
model and a fitting curve proportional to $|T_c -T|^{0.32}$, respectively.}
\end{figure}

Now we turn to study the magnetic fluctuation effects in the
paramagnetic state.  The magnetic susceptibility directly
measures the magnetic fluctuation as shown in the following equation:
\begin{eqnarray}
\chi\equiv \frac{\partial M }{\partial h} &=&
\frac{\partial}{L^3\partial h} \frac{{\rm Tr}{\sum_iS_i^z}
e^{-\beta({\cal H} - \mu {\cal N} -h\sum_iS_i^z)}}{{\rm
Tr}e^{-\beta({\cal H} - \mu {\cal N} -h\sum_iS_i^z)}} \nonumber\\ &=&
\frac{\langle M^2\rangle - \langle M\rangle^2}{T}.
\end{eqnarray}
Figure 2 provides $\chi$ as a function of $T/T_c$ denoted by open
squares.  With increasing temperature, $\chi$ increases to a maximum
near $T_c$ and then quickly decreases above $T_c$. It should be noted
that the susceptibility is well described by a power law of $\chi \sim
|T-T_c|^{-\gamma}$ as $T$ approaches close to the critical temperature
$T_c$ in the paramagnetic state, where the critical fluctuations are
dominant. The inset of Fig.~2 shows $\ln(\chi)$ vs. $\ln(|T-T_c|)$
curve, yielding the exponent $\gamma=0.92 \pm 0.01$ and the dashed
line in the Fig.~2 is a fitting curve. This value of the critical
exponent $\gamma$ is smaller than those of the experimental suggestion
of $1.22$ for La$_{0.7}$Sr$_{0.3}$MnO$_3$\cite{ghosh} and the
mean-field theory of Curie-Weiss law which is $\gamma=1$\cite{kittel}.
However, it is not unreasonable if we take account of the fact that
the present value is determined only from the $6^3$ finite size
cluster calculation with no scaling analysis.

\begin{figure}
\centering \leavevmode \epsfxsize=7.8cm \epsfbox{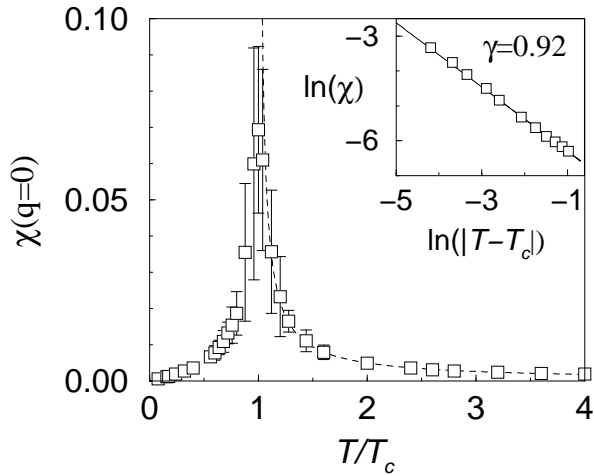}
\caption{Temperature dependence of magnetic susceptibility $\chi$. The
  dashed line is the fitting curve of $|T-T_c|^{-\gamma}$ with
  $\gamma=0.92$. The vertical lines indicate the estimation of error bar.
  Inset shows $\ln(\chi)$ vs. $\ln(|T-T_c|)$ above $T_c$. }
\end{figure}

On the other hand, in the high temperature regime of $\sim 2T_c$, the
magnetic susceptibility of the DE system exhibits a peculiar
temperature dependence.  In fact, these magnetic entities in the
paramagnetic state have been investigated over a long period of
time. For instance, $d(\chi^{-1})/dT$ in the high temperature
regime\cite{white} of a typical itinerant ferromagnetic such as nickel
($T_c=623$~K) is known to be temperature-dependent.  In order to
understand the spin entities in the paramagnetic state, we plot
$1/\chi$ vs. $T/T_c$ in Fig.~3. In the high-temperature region of $T
\agt 2.5T_c$, the corresponding $1/\chi$ follows a linear Curie-Weiss
behavior where each spin has a non-interacting magnetic moment. The
solid line is the fitting curve of the Curie-Weiss law, $\chi \approx
C/(T-\Theta)$, yielding the mean-field Curie temperature $\Theta=0.5
T_c$ and $C=0.0008$. However, starting from and below $2T_c$, $1/\chi$
shows a distinct deviation from the expected Curie-Weiss law,
suggesting a possible presence of short-range ferromagnetic clusters
due to the spin fluctuations above $T_c$. This feature are consistent
with observations in the recent experimental measurements on the
single crystal of La$_{0.7}$Ca$_{0.3}$MnO$_3$\cite{jaime}, the thin
film of La$_{0.6}$Y$_{0.07}$Ca$_{0.33}$MnO$_3$\cite{sun}, and the
layered material of La$_{1.35}$Sr$_{1.65}$Mn$_2$O$_7$ \cite{chauvet}
in low magnetic field $h < 0.1$~T.  For the high magnetic field,
however, $1/\chi$ becomes larger than the one expected from the
Curie-Weiss law\cite{sun,chauvet,oseroff}, indicating a suppression of
ferromagnetic fluctuation due to the carrier delocalization and
magnetic ordering by the applied magnetic.\cite{teresa,radaelli}

Assuming that ferromagnetic clusters exist, one should see an
effective moment which is larger than that of the appropriate average
of Mn$^{3+}$ and Mn$^{4+}$ moments. From the consideration that $P$
neighboring sites with ferromagnetically aligned spins form a spin
polaron of the spin $(S_1+PS_2)$ due to the double exchange mechanism,
the effective spin for the fluctuation effects becomes\cite{varma}
$$S_{\sf eff}^2=x(S_1+PS_2)(S_1+PS_2+1)+(1-x-Px)S_2(S_2+1),$$ where
$S_1$ and $S_2$ identify the spin of Mn$^{4+}$ and Mn$^{3+}$ species,
respectively. This estimation of the square of total spin will become
larger than the one without spin polaron formation, i.e., $P=0$ case,
and consequently the inverse susceptibility will be smaller compared
to the Curie-Weiss law.  These features are consistent with our
current Monte Carlo results as well as other results of the exact
diagonalization study\cite{horsch} in which the effective spins of
$e_g$ carriers grow from $1/2$ to 7 and the EPR experiments above
$T_c$.\cite{chauvet} Thus, it is reasonable to interpret that the spin
polarized carriers form ferromagnetic polarons with each individual
cluster retaining a large magnetic moment.

\begin{figure} 
\centering \leavevmode \epsfxsize=7.8cm \epsfbox{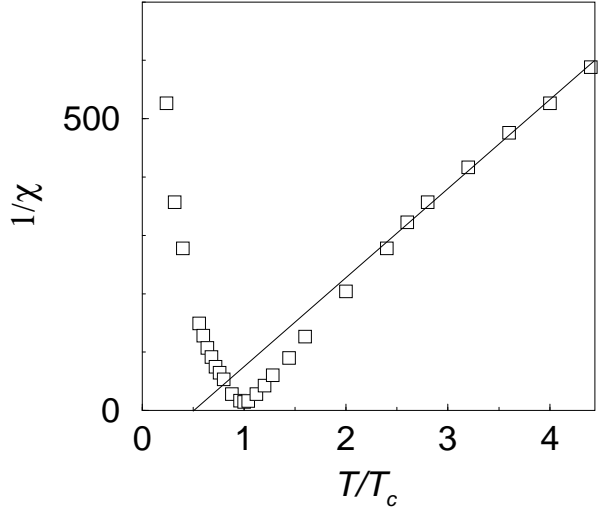}
\caption{Temperature dependence of the inverse susceptibility
$1/\chi$. The solid line is the best fit of the Curie-Weiss behavior
of the form $\chi=C/(T-0.5T_c)$.}
\end{figure}

It is interesting to compare our Monte Carlo results with the outcomes
from the EPR measurements. For this purpose we define the deviation of
susceptibility from the Curie-Weiss law as $\Delta\chi\equiv
\chi-C/(T-\Theta)$. In the intermediate temperature region of $T_c < T
\alt 2.5T_c$, we find that $\Delta \chi$ can be fit by the following
form
\begin{equation}
\Delta \chi = \chi_o\exp(E_a/T)
\end{equation}
where $E_a$ is an activation energy which scales to the $T_c$. Figure
4 shows $\ln(\Delta \chi)$ vs. $1000/T$ for fixing $T_c=270$~K denoted
open squares. The solid line is a fitting curve, yielding $E_a=10T_c$
which corresponds to $0.22$~eV for $t=0.17$~eV, when
$\chi_o=9\times10^{-6}$ is used.

It is interesting to note that the EPR intensity decreases
exponentially above $T_c$ and has a strong correlation with the
deviation of the magnetic susceptibility\cite{chauvet,oseroff}. For a
comparison with our result of $\Delta\chi$, the log of intensity as a
function of $1000/T$ for La$_{2/3}$Ca$_{1/3}$MnO$_3$ ($T_c \approx
270$~K) from Ref.~[18] is shown in the inset of the Fig.~4. The shape
of the $T$-dependence of EPR intensity is quite similar to the Monte
Carlo result of $\Delta\chi$. The $E_a$ obtained from EPR measurement
is approximately $7T_c$ which is slightly smaller than that obtained
from the $\Delta\chi$. According to the studies of Oseroff {\it et
al.},\cite{oseroff} those spin entities are associated with some form
of the magnetic cluster with the spins of about $\sim 30$ per formula
unit. In other words, the $E_a$ would be required in order to
dissociate these spin entities made of collection of individual spins.
Therefore, it is suggested that the magnetic polaron formation driven
by the spin-disorder scattering should be generic of the MI transition
regimes and is likely associated with the CMR phenomena in the doped
manganites.

\begin{figure} 
\centering \leavevmode \epsfxsize=7.8cm \epsfbox{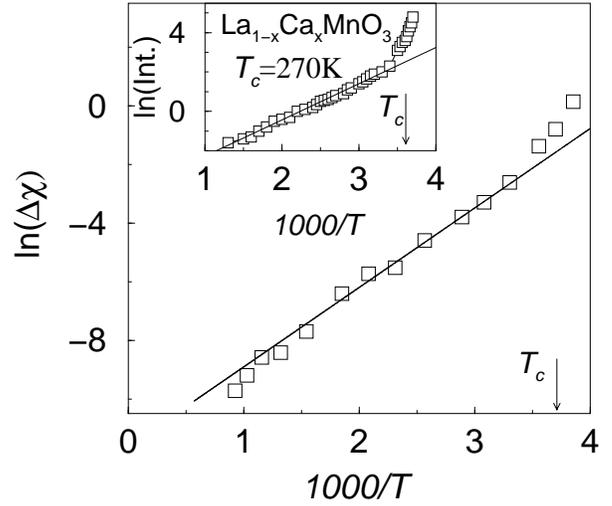}
\caption{ $\ln(\Delta \chi)$ vs. $1000/T$ for $T_c\approx 270$~K and
the solid line represents the best fit to Eq.~(6). In the inset, EPR intensity
of $\ln(I)$ vs. $1000/T$ for La$_{2/3}$Ca$_{1/3}$MnO$_3$ ($T_c \approx
270$~K) from Ref.~[18].}
\end{figure}

{}Finally, we estimate the effective sizes of the ferromagnetic
clusters in the paramagnetic state. Figure~5 shows the spin-spin
correlation $\langle \sum_i\vec{S}_i\cdot\vec{S}_{i+r}\rangle$ as a
function of $T/T_c$ in the vicinity of $T_c$ for several values of
lattice sites. The shortest correlation for $r=1$ is robust even for
$T\agt 2T_c$ and, in particular, almost $15\%$ of the maximum at
$T_c$. Furthermore, the short-range ferromagnetic correlation is
clearly seen in the intermediated regime of $T_c < T \alt 2T_c$. The
size of the ferromagnetic surroundings is of the order of $3\sim 4$
lattice spacing distances, i.e., 12\AA $\sim$16\AA\ if we consider a
size of lattice as $4$\AA, which is in good agreement with recent
neutron scattering.\cite{teresa,radaelli} Moreover, a simple
estimation of the ferromagnetic correlation length, $P
\approx(at/T)^{3/5}$, in terms of a ferromagnetic spin-polaron picture
has been introduced by Varma\cite{varma}.  At the temperature range of
paramagnetic state, in particular at room temperature $\sim 300$~K,
the size of the spin polaron is estimated to be a few lattice size for
$t\approx 0.2$~eV, which agrees with the current results.  It implies
that the spin-polarized carriers can be trapped into a local
ferromagnetic surroundings due to the spatially fluctuating spin
correlations, resulting in the formation of the magnetic polarons in
the paramagnetic phase.\cite{varma,horsch}

\begin{figure} \centering
\leavevmode \epsfxsize=7.8cm \epsfbox{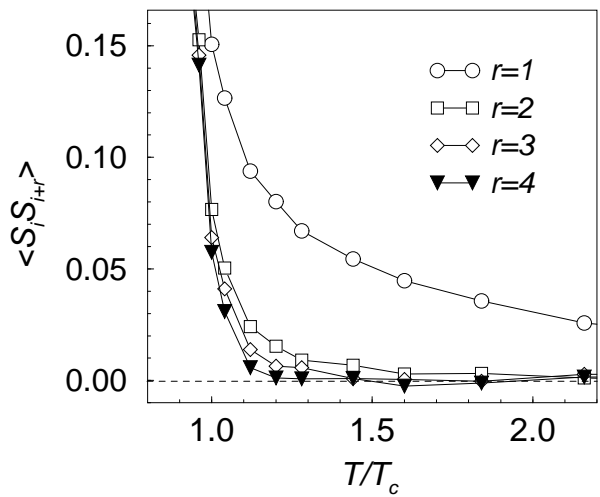}
\caption{Temperature dependence of the spin correlation $\langle \sum_i\vec{S_i}\cdot\vec{S}_{i+r}\rangle$ for $x=0.5$
and $L^3=6^3$.}
\end{figure}

\section{Conclusions}

Using Monte Carlo methods, we have studied the magnetic fluctuation
effects near and above $T_c$ within the framework of DE model. We
found that the temperature dependence of the susceptibility shows a
sharp peak near $T_c$, and the inverse susceptibility displays
non-Curie-Weiss behavior above $T_c$. The activation energy obtained
from the deviation of susceptibility from the Curie-Weiss behavior is
consistent with the EPR measurements.  These results clearly
demonstrate the formation of the magnetic polaron with short-range
ferromagnetic ordering in the paramagnetic state.  Moreover, the
ferromagnetic correlation length is estimated to be $3\sim 4$ lattice
spacings which is in good agreement with recent neutron scattering
experiments\cite{teresa,radaelli}. From the results, it is suggested
that the magnetic polaron formation is responsible for the magnetic
transition and the magneto-transport properties in doped CMR
manganites.

\acknowledgements 

This work was supported by the Creative Research Initiatives Program
in Korea. J.Y. acknowledges the support of the KOSEF and the BSRI
(015-D00118) program. The author gratefully acknowledges the resource
of the ETRI supercomputer center used in the present work.

\end{document}